\documentclass[aps,twocolumn,amsmath,amssymb,showpacs,superscriptaddress,notitlepage]{revtex4-2}
\usepackage[T1]{fontenc}
\usepackage[utf8]{inputenc}
\usepackage{color}

\makeatletter
\usepackage{amssymb}
\usepackage{amsmath}
\usepackage{graphicx}
\usepackage[colorlinks=true,linkcolor=blue,anchorcolor=red,citecolor=blue,urlcolor=blue]{hyperref}
\usepackage[caption=false]{subfig}
\usepackage{lipsum}

\makeatother

\usepackage{babel}

\begin{document}
%linenumber
%\linenumbers	
	
\renewcommand{\figurename}{Figure}

\title{Intertwined Weyl phases emergent from higher-order topology and unconventional Weyl fermions via crystalline symmetry}
\author{W.\ B. Rui}
\email{wbrui@hku.hk}

\address{Department of Physics and HKU-UCAS Joint Institute for Theoretical
and Computational Physics at Hong Kong, The University of Hong Kong,
Pokfulam Road, Hong Kong, China}

\author{Zhen Zheng}
\address{Department of Physics and HKU-UCAS Joint Institute for Theoretical
	and Computational Physics at Hong Kong, The University of Hong Kong,
	Pokfulam Road, Hong Kong, China}

\author{Moritz M. Hirschmann}
\address{Max Planck Institute for Solid State Research, Heisenbergstrasse 1,
	D-70569 Stuttgart, Germany}

\author{Song-Bo Zhang}
\address{Institute for Theoretical Physics and Astrophysics, University of
	W\"urzburg, D-97074 W\"urzburg, Germany}

\author{Chenjie Wang}
\email{cjwang@hku.hk}

\address{Department of Physics and HKU-UCAS Joint Institute for Theoretical
	and Computational Physics at Hong Kong, The University of Hong Kong,
	Pokfulam Road, Hong Kong, China}

\author{Z.\ D. Wang}
\email{zwang@hku.hk}

\address{Department of Physics and HKU-UCAS Joint Institute for Theoretical
	and Computational Physics at Hong Kong, The University of Hong Kong,
	Pokfulam Road, Hong Kong, China}

\date{\today}
\begin{abstract}
We discover three-dimensional intertwined Weyl phases, by developing a theory to create topological phases. The theory is based on intertwining existing topological gapped and gapless phases protected by the same crystalline symmetry. The intertwined Weyl phases feature both unconventional Weyl semimetallic (monopole charge>1) and higher-order topological phases, and more importantly, their exotic intertwining. While the two phases are independently stabilized by the same symmetry, their intertwining results in the specific distribution of them in the bulk. The construction mechanism allows us to combine different kinds of unconventional Weyl semimetallic and higher-order topological phases to generate distinct phases. Remarkably, on 2D surfaces, the intertwining causes the Fermi-arc topology to change in a periodic pattern against surface orientation. This feature provides a characteristic and feasible signature to probe the intertwining Weyl phases. Moreover, we provide guidelines for searching candidate materials, and elaborate on emulating the intertwined double-Weyl phase in cold-atom experiments.
\end{abstract}

\maketitle
\noindent\textbf{INTRODUCTION}\quad\\
Owing to fertile ground of crystalline symmetries, topological gapless phases,
characterized by nontrivial band degeneracies, have been undergoing rapid development in condensed matter physics~\citep{Chiu_RMP_2016,Armitage2018,Yang_PRL_2014}.
A representative example is the diversification of prototypical Weyl phases. Beyond conventional Weyl fermions with monopole charge $\pm1$~\citep{Wan_PRB_2011,xu_discovery_2015,Lv_Weyl_2015,Meng_PRB_2012,Morali_magnetic_Weyl_2019,Liu_magnetic_Weyl_2019}, unconventional Weyl
fermions, which possess a higher monopole charge due to crystalline symmetry, were discovered~\citep{Bradlynaaf5037,sanchez_topological_2019,yang_topological_2019,he_observation_2020,Fang2012PRL,chen_photonic_2016,huang_new_2016,Vaidya_2020_charge2,Yang2020PRL,Dantas2020PRR}.
For instance, threefold Weyl fermions with charge $\pm2$ can be stabilized
by a nonsymmorphic symmetry~\citep{Bradlynaaf5037,sanchez_topological_2019,yang_topological_2019,he_observation_2020},
and double(triple)-Weyl fermions with a quadratic(cubic) twofold degeneracy
can be protected by a rotation symmetry~\citep{Fang2012PRL,chen_photonic_2016,huang_new_2016,Vaidya_2020_charge2,Yang2020PRL,Dantas2020PRR}.

On the other hand, crystalline symmetries have largely enriched
topological gapped phases~\citep{Liang2011crystalline,hsieh_topological_2012,tanaka_experimental_2012,dziawa_topological_2012,Ando_Topological_2015,Fang_2015_new,Slager_RPX_2017,Khalaf_2018_symmetry,po_symmetry-based_2017,tang_comprehensive_2019,zhang_catalogue_2019,vergniory_complete_2019}.
Among these phases, higher-order topological insulators, which feature anomalous 
boundary states, have drawn particular attention recently~\citep{Benalcazar61,song_PRL_2017,Piet_PRL_2017,Peterson_fractional_2020,Schindl18science,Schindler2018higher,serra2018observation,Ezawa18PRL,Bernevig2019PRL,ChenR20PRL,ZhangSB20arxiv2,Trifunovic19PRX,Eslam_PRB_2018,ZhangSB20quantumcomputation}.
In contrast
to conventional bulk-boundary correspondence, i.e., a $d$-dimensional
bulk topology corresponds to $(d-1)$-dimensional boundary states,
the boundary states of higher-order topological insulators are further
restricted by crystalline symmetries and exhibit boundary states in
a lower dimension, e.g., corner or hinge states. Recently, it has been found that higher-order topological
semimetals (e.g. Dirac, Weyl, and nodal-line)
can be realized by nontrivially stacking higher-order insulators~\citep{Lin_PRB_2018,wang2020boundary,ghorashi_second-order_2019,Tiwari2020PRR,Bitan2019PRR,wang_2020_higherorder,ghorashi_2020_higherorder,rui_higher-order_2020,wei_higher-order_2021,luo_observation_2021,zhang_2019_higherorder,Bitan-general-principle,Bitan-dirty-higher-order}.
Specifically, higher-order Weyl semimetals were discovered by stacking two-dimensional (2D) higher-order topological insulators, with broken time-reversal and/or inversion symmetry, in a third dimension~\citep{Bitan2019PRR,wang_2020_higherorder,ghorashi_2020_higherorder,rui_higher-order_2020,wei_higher-order_2021,luo_observation_2021}.
Note that in this process, for electronic systems, the Kramers degeneracy of the original higher-order topological insulators, if any, has to be broken before stacking~\citep{Bitan2019PRR}.
In this regard, the higher-order Weyl semimetal phase depends on the 2D sub higher-order phases.  
The phase transition between these phases results in the higher-order Weyl points, which are typically of conventional type with charge $\pm1$.

Though topological gapless phases characterized by nontrivial band degeneracies, and topological gapped phases featuring anomalous boundary states, can be stabilized by the same crystalline symmetry, so far, their interplay remains to be explored.

In this work, we develop a theory to generate topological phases of matter, by intertwining existing ones protected by a crystalline symmetry. 
We discover intertwined Weyl phases, which feature the exotic interplay of unconventional Weyl fermions and higher-order topology. In the bulk, the two phases are independently stabilized by the crystalline symmetry:
the Weyl fermions have a monopole charge larger than 1 by the symmetry, and a higher-order topological phase was further
superposed by the symmetry. The intertwining results in a specific distribution
of the two phases in the bulk, i.e., the higher-order topology exists in the region outside pairs of unconventional Weyl points with opposite charges. Due to their independence, the two phases are separately tunable. The combination of different unconventional Weyl semimetallic and higher-order topological phases results in distinct intertwined Weyl phases. Note that this mechanism is different from the higher-order Weyl semimetal which is determined by the higher-order topology as discussed above.

The intertwined Weyl phases are characterized by the the drastic change of Fermi-arc topology on 2D surfaces upon rotating surface termination. This phenomenon comes from the intertwining between the two constituent phases. Even though higher-order topological phases feature boundary states on 1D hinges, they can intertwine with the unconventional Weyl phase on 2D 
surfaces of the system: Fermi arcs form due to the underlying unconventional Weyl
phases, while the topology of these Fermi arcs is drastically changed by the higher-order topological phase.
Thus, we identify a prominent topological feature of the intertwined phase: the change of Fermi-arc topology against surface orientation in a periodic pattern. 
The period is determined by rotation symmetry, e.g., a $2n$-fold
rotation symmetry leads to a period of $\pi/n$. Specifically,
we discuss intertwined double-Weyl phases, where double-Weyl fermions
and higher-order topology are intertwined due to a fourfold rotation symmetry.
The topological phase is characterized by the periodic change of Fermi-arc 
topology with period $\pi/2$. The periodic behavior can be perfectly explained
by an effective boundary Hamiltonian, in accordance with our theory.
Finally, we show that intertwined double-Weyl phases are realizable in cold-atom
experiments, which can serve as a promising platform to implement our theory in experiment.

\begin{figure}
	\includegraphics[width=0.9\columnwidth]{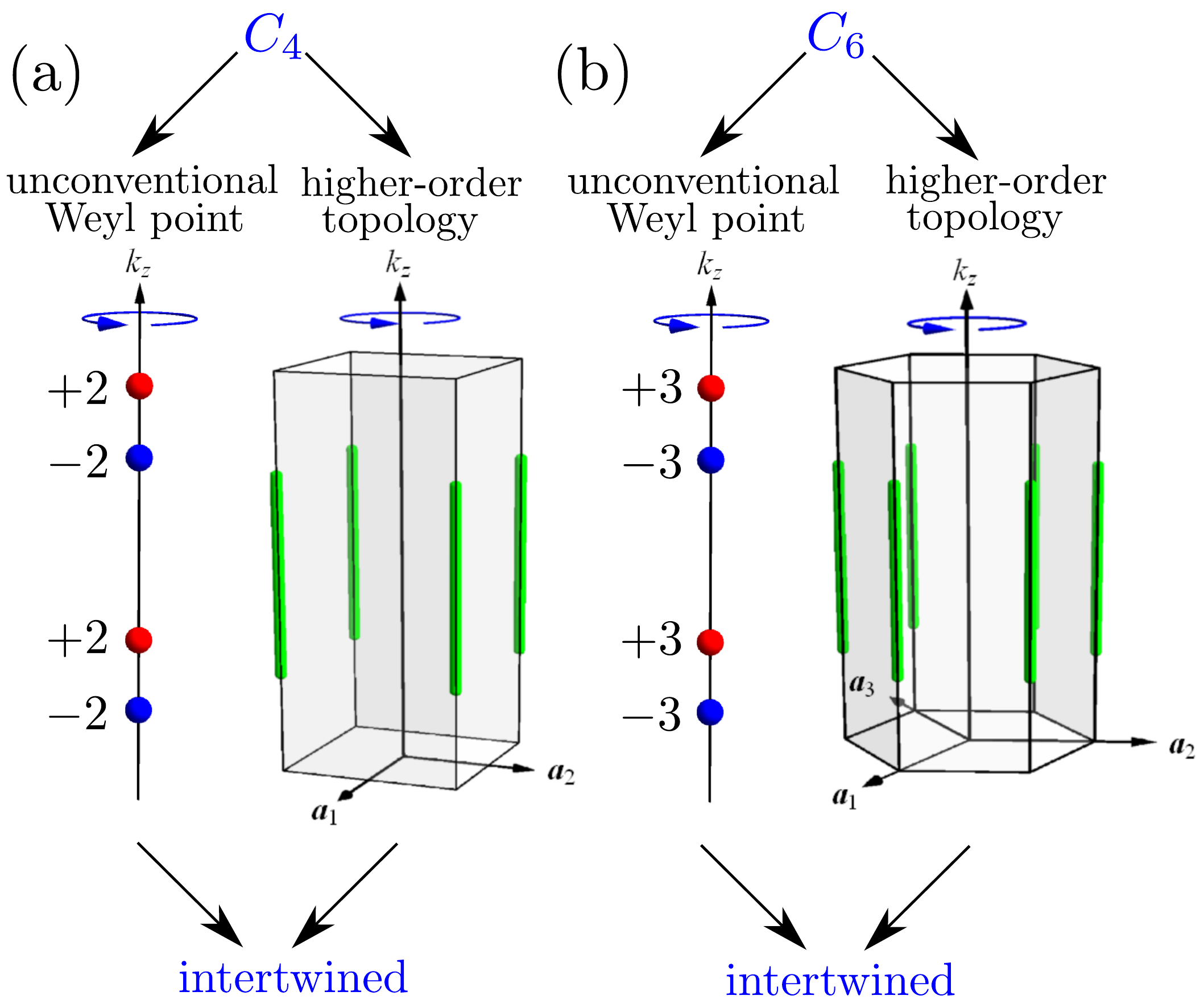}
	\caption{\textbf{Intertwining topological phases by crystalline symmetry.} Red and blue dots represent Weyl points, and green sold lines denote hinge states from higher-order topology.
		(a) The fourfold rotation symmetry ($C_4$) stabilizes double-Weyl points and higher-order topology at the same time. The interplay between the two 
		phases results in an intertwined double-Weyl phase.
		(b) The sixfold rotation symmetry ($C_6$) stabilizes triple-Weyl points and higher-order topology at the same time, resulting in an intertwined triple-Weyl phase. \label{fig:intertwine}}
\end{figure}

\vspace{0.5cm}
\noindent\textbf{RESULTS}\quad\\
\noindent\textbf{Intertwining topological phases by crystalline symmetry}\quad\\
We begin by showing
how to intertwine topological phases by crystalline symmetry.
The essential physics can be captured by the following simple but
generic Hamiltonian in 3D momentum space,
\begin{equation}
	\begin{aligned}
		H(\mathbf{k}) & =H_{\text{Weyl}}+m\Gamma_{\text{h}} \\
		& =k_{+}^{n}\tau_{3}\sigma_{-}+k_{-}^{n}\tau_{3}\sigma_{+}+k_{z}\tau_{3}\sigma_{3}+m\Gamma_{\text{h}},
	\end{aligned}\label{eq:hal-main}
\end{equation}
where $k_{\pm}=k_{x}\pm ik_{y}$, $\sigma_{\pm}=(\sigma_{1}\pm i\sigma_{2})/2$,
and $\sigma_{i}$ and $\tau_{i}$ ($i=1,2,3$) are Pauli matrices
for (pseudo)spin and orbital degrees of freedom, respectively.
$H_{\text{Weyl}}$, responsible for generating unconventional Weyl
points, is superposed with $m\Gamma_{\text{h}}$, a term for introducing higher-order
topology. $m$ is a real model parameter, and $\Gamma_{\text{h}}$ a $4\times4$
matrix. 
After being projected to the boundary, 
$m\Gamma_{\text{h}}$ acts as a mass term that is constrained by the symmetry, as we will discuss in Eqs.~\eqref{eq:boundary-hal} and~\eqref{eq:antiperiodic}.
Each of the two parts is invariant under the $2n$-fold rotation
symmetry $C_{2n}$ about the $z$ axis, with $2n\leq6$ by lattice
restriction ($n\in\{2,3\}$).

\begin{figure}
	\includegraphics[width=0.95\columnwidth]{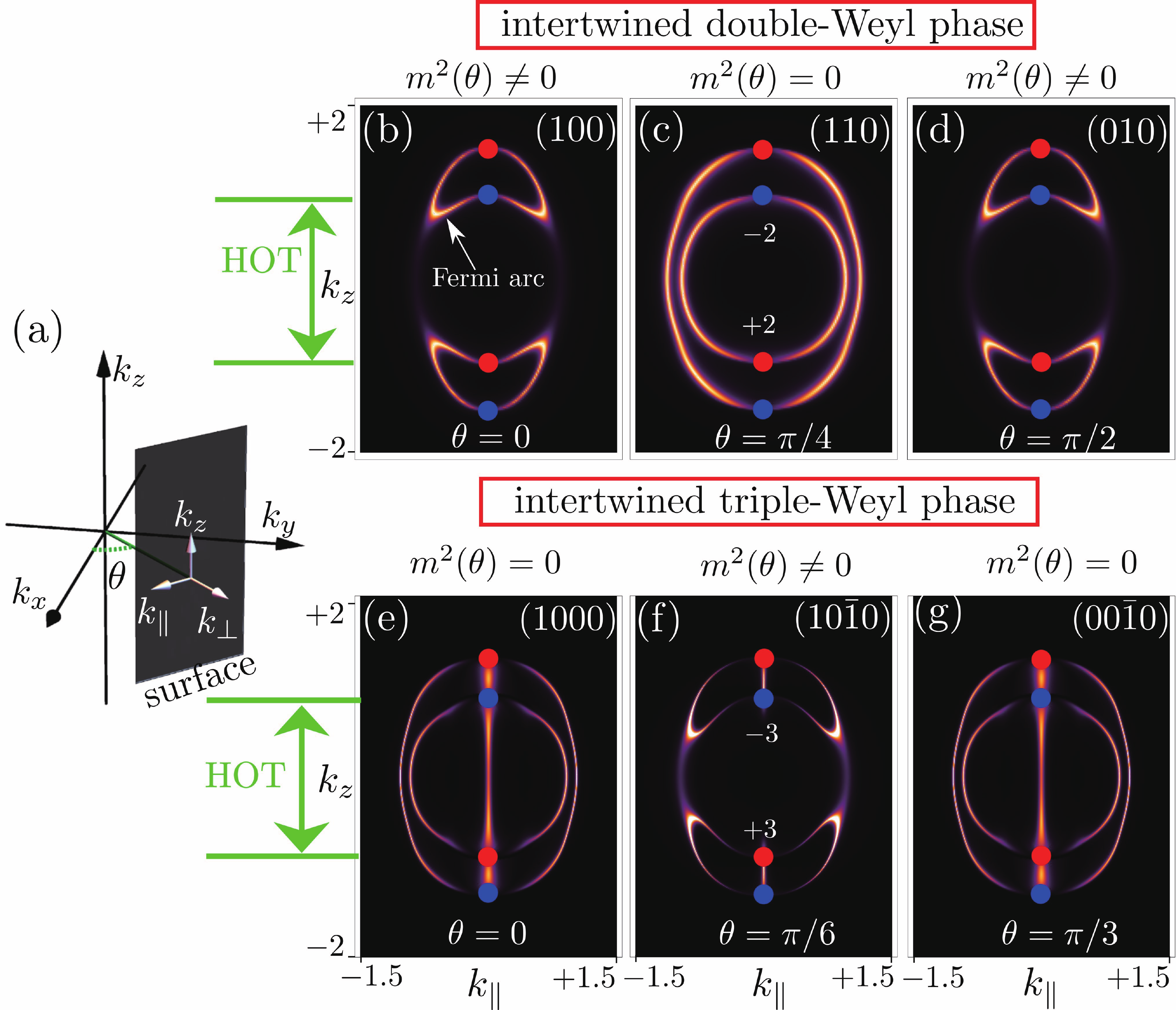}
	\caption{\textbf{The intertwining results in the change of Fermi-arc topology in a periodic pattern.} (a) The relation between different coordinate systems. $\theta$ indicates the orientation of the surface (black
		plane), on which the local density of states is computed numerically.
		(b-d) Fermi arcs of intertwined double-Weyl phase in a full period
		of $\theta\in[0,\pi/2]$. Each projected double-Weyl point emits two Fermi arcs on the surface.
		The topology of Fermi arcs changes when rotating from the $(100)$
		surface ($\theta=0$) to the $(110)$ surface ($\theta=\pi/4$), and
		it returns to the initial topology after rotating further to the
		$(010)$ surface ($\theta=\pi/2$), completing one period. (e-g) Fermi
		arcs of intertwined triple-Weyl phase in a full period of $\theta\in[0,\pi/3]$.
		Each projected triple-Weyl point emits three Fermi arcs on the surface. The topology of
		Fermi arcs changes when rotating from the $(1000)$ surface ($\theta=0$)
		to the $(10\bar{1}0)$ surface ($\theta=\pi/6$), and it returns to
		the initial topology after rotating further to the $(00\bar{1}0)$
		surface ($\theta=\pi/3$), completing one period. \label{fig:periodic_arcs}}
\end{figure}

As shown in Figure~\ref{fig:intertwine}, the intertwining can be induced by a crystalline symmetry that superposes the two phases 
at the same time:
\begin{enumerate}
\item The bulk Weyl points with higher monopole charge are stabilized by rotation symmetry.
Unconventional Weyl points with charge $\pm n$ ($n>1$)
can be generated by breaking time-reversal and/or inversion
symmetry in $H_{\text{Weyl}}$. After the substitution of $(k_{x},k_{y})=(k\cos\theta,k\sin\theta$)
{[}see Figure~\ref{fig:periodic_arcs}(a){]}, $H_{\text{Weyl}}$ can
be rewritten as $H_{\text{Weyl}}(\theta)=k^{n}e^{+in\theta}\tau_{3}\sigma_{-}+k^{n}e^{-in\theta}\tau_{3}\sigma_{+}+k_{z}\tau_{3}\sigma_{3}$.
As can be seen, these unconventional Weyl points are protected by
rotation symmetry \citep{Fang2012PRL}
\begin{equation}
\hat{C}_{2n}H_{\text{Weyl}}(\theta)\hat{C}_{2n}^{-1}=H_{\text{Weyl}}(R_{2n}\theta).
\end{equation}
Here, $\hat{C}_{2n}=\tau_{3}\sigma_{3}$, and $R_{2n}\theta=\theta+\pi/n$
acts in momentum space.
\item The higher-order topology is further enforced by rotation symmetry. The topological protection by symmetry can be explicitly
demonstrated at the boundary. After projecting the Hamiltonian to
the boundary subspace, we can get an effective Hamiltonian on the
surface in the form of (See Supplementary Note 2 for derivation details.)
\begin{equation}
h_{\text{surface}}=\sum_{i=0}^{n}a_{i}k_{\parallel}^{i}\sigma_{3}+m(\theta)\gamma_{\text{h}},\label{eq:boundary-hal}
\end{equation}
where $a_{i}$ is a real coefficient, which may depend on $k_{z}$,
$\theta$ is the surface orientation, and $k_{\parallel}$ the momentum
parallel to the surface {[}see Figure~\ref{fig:periodic_arcs}(a){]}.
The higher-order term $m(\theta)\gamma_{\text{h}}$, originated
from $m\Gamma_{\text{h}}$, is still rotation-symmetric, $\hat{c}_{2n}m(\mathbf{\theta})\gamma_{\text{h}}\hat{c}_{2n}^{-1}=m(R_{2n}\mathbf{\theta})\gamma_{\text{h}}$,
with the projected rotation operator $\hat{c}_{2n}=\sigma_{3}$. 
Note that $m(\theta)$, which stems from the constant $m$ term in Eq.~\eqref{eq:hal-main}, becomes $\theta$-dependent
after the projection by the orientation-dependent boundary wavefunctions. $\gamma_{\text{h}}$
anticommutes with $\sigma_{3}$, e.g., $\sigma_1$ or $\sigma_2$, so that it acts as a mass
term in Eq.~(\ref{eq:boundary-hal}). Thus, we obtain
\begin{equation}
m(\mathbf{\theta})=-m(R_{2n}\mathbf{\theta})=-m(\theta+\pi/n).\label{eq:antiperiodic}
\end{equation}
This relation means $m(\mathbf{\theta})$ must have a zero value between $\theta$
and $\theta+\pi/n$, which leads to a gapless point in the surface spectrum.
The 1D Hamiltonian 
(\ref{eq:boundary-hal}) for a fixed $k_z$ is characterized by a $\mathbb{Z}$ topological invariant. 
Note that the same mechanism leads to higher-order topological
insulators protected by crystalline symmetry, where gapless points correspond to corner or hinge modes~\citep{Trifunovic19PRX,Eslam_PRB_2018}. 
Here, as a manifestation of higher-order topology, the gapless points 
result in 1D hinge states, as shown by the green solid lines in Figure~\ref{fig:intertwine}.

\end{enumerate}

The two topological phases in superposition no longer act 
individually, but intertwine with each other, as we discuss next. Thus, we refer to the resultant phase as ``intertwined Weyl phase'', which is different from each individual phase.

We emphasize that in the intertwined Weyl phase, the
unconventional Weyl points and the higher-order topology are independently stabilized  by the symmetry. This is different from 
the higher-order Weyl semimetals, where the higher-order Weyl points depend on the higher-order topological phases and are typically conventional ones of topological charge-1~\citep{wang_2020_higherorder,ghorashi_2020_higherorder,rui_higher-order_2020,wei_higher-order_2021,luo_observation_2021}.
Due to the independent nature of the two phases, 
they can be tuned separately to generate various intertwined Weyl phases. Specifically, different unconventional Weyl fermion phases can be incorporated to build
with higher-order topological phase under the same or different crystalline symmetries, resulting in
distinct kinds of intertwined Weyl phases, as we will discuss later.

\vspace{0.5cm}
\noindent\textbf{Intertwining and the change of Fermi-arc topology in a periodic pattern}\quad\\
In the bulk, the intertwining results in a specific
distribution of the two phases. This is because the higher-order topology cannot survive in regions with non-trivial Chern number. To have vanishing Chern number, 
it is required that the unconventional Weyl points with opposite charges appear in pairs. In this way, there can exist regions with trivial Chern numbers to host
higher-order topology. As shown in Figure~\ref{fig:periodic_arcs} by green arrows, the higher-order topological (HOT) phase exists outside pairs of unconventional Weyl points with opposite charges.

Inside pairs of unconventional Weyl points, e.g., the lower and upper two sets of Weyl points in Figure~\ref{fig:periodic_arcs}(b), there is no $m(\theta)$, the surface Hamiltonian (\ref{eq:boundary-hal})
describes Fermi arcs like those in unconventional Weyl semimetals. On 2D surfaces, the existence of these Fermi arcs is isotropic, i.e., they do not depend on surface orientation
$\theta$. However, in HOT regions, the higher-order topological term $m(\theta)$ exists. The existence of Fermi arcs in this region depends on $m(\theta)$ and becomes anisotropic. We find that due to the intertwining of the two parts of Fermi arcs, the Fermi-arc topology on 2D surfaces is drastically changed. We use $m(\theta)$ in Eq.~\eqref{eq:antiperiodic} to explain the phenomenon in detail below.

First, the gapless point enforced by the higher-order
topological phase in Eq.~\eqref{eq:antiperiodic} crucially affects the Fermi-arc topology.
We use the spectrum of the surface Hamiltonian~\eqref{eq:boundary-hal}, $E_{\text{surface}}=\pm\sqrt{(\sum_{i=0}^{n}a_{i}k_{\parallel}^{i})^2+m^{2}(\mathbf{\theta})}$, for illustration.
If there is no gapless point [$m(\mathbf{\theta})\neq0$] in the HOT phase region for a specific angle $\theta$, no Fermi arc can go through this HOT phase region because $E_{\text{surface}}$ is gapped, e.g., as shown by Figures~\ref{fig:periodic_arcs} (b) and (d). In contrast, if there is a gapless point [$m(\mathbf{\theta})=0$] in HOT region, Fermi arcs can go through this HOT region because $E_{\text{surface}}$ is gapless for some $k_\parallel$,  e.g., as shown by Figure~\ref{fig:periodic_arcs} (c).
In this regard, the gapless point [$m(\mathbf{\theta})=0$] determines whether the 
Fermi arcs can go through the higher-order phase region at an angle $\theta$. Thus, strikingly, the Fermi-arc topology changes with $\theta$.

Second, $m(\theta)$ in Eq.~(\ref{eq:antiperiodic}) is antiperiodic. Remarkably,
it becomes periodic in the surface spectrum $E_{\text{surface}}$ after taking the square, since the square of $m(\theta)$ obeys
\begin{equation}
m^{2}(\mathbf{\theta})=m^{2}(\theta+\pi/n),\label{eq:mass-period}
\end{equation}
which has a period of $\pi/n$. 

The periodic function $m^{2}(\theta)$
indicates that the topology of Fermi arcs not only changes with the surface orientation $\theta$, but also in a periodic pattern. 
We emphasize that the change of Fermi-arc topology comes from the intertwining of 
unconventional Weyl fermions and higher-order topology. Thus, this phenomenon
is absent in any individual phase alone. This is distinct from Ref.~\cite{zhang_second-order_2019}, where the edge states in different samples of different geometry are
determined by higher-order topology. Thus, it constitutes the
characteristic feature of intertwined Weyl phases.

After figuring out how $m(\theta)$ influences the behaviour of boundary states on the 2D surfaces, 
it is also necessary to discuss its impact on the boundary states on the 1D hinges.
The hinge states, seen in the HOT phases, are determined by the condition $m(\theta)=0$.
For $m(\theta)\neq0$, no localized zero modes can be observed. This is in accordance with the 
behaviour of the higher-order topological phase. Thus, the hinge states of higher-order topology can still
be observed in 1D, as shown by the green lines in Figure~\ref{fig:intertwine}.

\vspace{0.5cm}
\noindent\textbf{ Exemplification of intertwined double-Weyl and triple-Weyl phases}\quad\\
The intertwined double-Weyl phase is generated by the fourfold rotation
symmetry $C_{4}$, i.e., $n=2$, as we show previously. The Weyl points carry a monopole charge
of $\pm2$, which emit two Fermi arcs on the surface. According to
Eq.~(\ref{eq:mass-period}), on 2D surfaces, the change of Fermi-arc topology is of period $\pi/2$.
In Figures~\ref{fig:periodic_arcs}(b, c, d), we show the topology
of Fermi arcs in three representative orientations of $\theta=0$,
$\theta=\pi/4$, and $\theta=\pi/2$ in a full period of $[0,\pi/2]$,
respectively. Clearly, by rotating surfaces, the Fermi-arc topology changes
from Figure~\ref{fig:periodic_arcs}(b) ($\theta=0$) to Figure~\ref{fig:periodic_arcs}(c)
($\theta=\pi/4$), and then returns to the initial topology in Figure~\ref{fig:periodic_arcs}(d)
($\theta=\pi/2$). On 1D  hinges, the boundary states occur at $m^2(\theta)=0$,
e.g., $\theta=n\pi/4$ with $(n=1,3,5,7)$. This is
in accordance with the boundary states of higher-order topological phase with $C_4$ symmetry, as shown in Figure~\ref{fig:intertwine}(a).

A similar story applies to the intertwined triple-Weyl phase which
is enforced by sixfold rotation symmetry $C_{6}$, i.e., $n=3$.
On 2D surfaces, each triple-Weyl point emits three Fermi arcs on the surface, and
the topology of Fermi arcs changes periodically with a period
of $\pi/3$, as shown by Figures~\ref{fig:periodic_arcs}(e, f, g)
in a full period $[0,\pi/3]$. On 1D  hinges, the boundary states occur at, e.g., $\theta=n\pi/6$ with $(n=0,2,4,6,8,10)$. They are in accordance with the boundary states of higher-order topological phase with $C_6$ symmetry, as shown in Figure~\ref{fig:intertwine}(b).

\vspace{0.5cm}
\noindent\textbf{Intertwined double-Weyl phase}\quad\\
We now apply our theory to the intertwined double-Weyl semimetals, which
can be realized in cold-atom experiments or other artificial systems. The model Hamiltonian reads
\begin{equation}
\begin{aligned}
H(\mathbf{k}) & =2A(\cos k_{y}-\cos k_{x})\tau_{3}\sigma_{1}+2A\sin k_{x}\sin k_{y}\tau_{3}\sigma_{2} \\
& +M(\mathbf{k})\tau_{3}\sigma_{3}+\epsilon\tau_{0}\sigma_{3}+m\tau_{1}\sigma_{1},
\end{aligned}\label{eq:lattice-model}
\end{equation}
where $M(\mathbf{k})=M_{0}-2t(\cos k_{x}+\cos k_{y}+\cos k_{z})$.
The four double-Weyl points are located on the $k_{z}$ axis
at $k_{z}=\pm k_{\text{w}1}$ and $k_{z}=\pm k_{\text{w}2}$, where $k_{\text{w}1(2)}=\arccos[(M-(+)\sqrt{\epsilon^{2}+m^{2}})/2t]$
with $M=M_{0}-4t$. By series expansion around the Weyl points at
$(0,0,\pm k_{\text{w}1(2)})$, i.e., $(k_{x},k_{y},k_{z})\rightarrow(\delta k\cos\theta,\delta k\sin\theta,\delta k_{z})$,
the form of the low energy model $H(\mathbf{k})=A\delta ke^{+2i\theta}\tau_{3}\sigma_{-}+A\delta ke^{-2i\theta}\tau_{3}\sigma_{+}+t\delta k_{z}\tau_{3}\sigma_{3}+\epsilon\tau_{0}\sigma_{3}+m\tau_{1}\sigma_{1}$
is the same as Eq.~(\ref{eq:hal-main}). Double-Weyl points are generated
by the $\epsilon$ term that breaks time-reversal symmetry. They possess monopole charge of $\pm2$~\cite{Bitan-dirty-weyl,Bitan_nonabelian_PRR}. $m\tau_{1}\sigma_{1}$
corresponds to $m\Gamma_{\text{h}}$ in Eq.~(\ref{eq:hal-main}), which is responsible
for the higher-order topology. The whole system is protected by fourfold
rotation symmetry $\hat{C}_{4}=\sigma_{3}\tau_{3}$. The Fermi arcs shown in Figures~\ref{fig:periodic_arcs} (b)-(d) are numerically calculated by using the model~\eqref{eq:lattice-model} with open boundary conditions.

\begin{figure}
	\includegraphics[width=0.9\columnwidth]{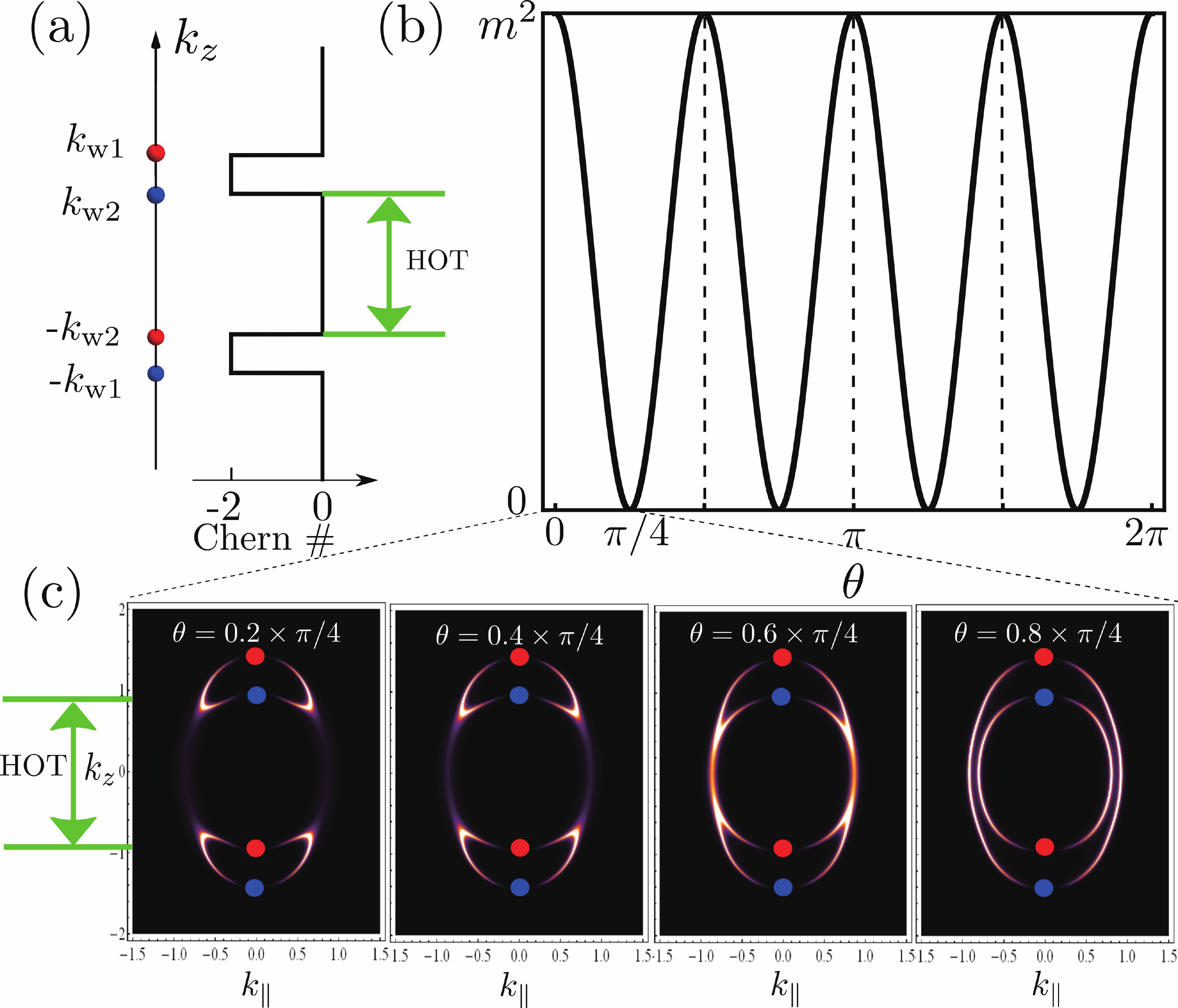}
	\caption{ \textbf{The location of Weyl points and the orientation dependent mass term.} (a) Left: The four double-Weyl points on the $k_{z}$ axis. Red and
		blue dots denote Weyl points with monopole charge $+2$ and $-2$,
		respectively. Right: Chern number on $k_{x}k_{y}$-plane against $k_{z}$,
		which changes by $\pm2$ when a double-Weyl point is passed.
		It is nontrivial within upper and lower two pairs of double-Weyl points.
		(b) The higher-order term $m(\theta)^{2}=m^{2}\cos^{2}(2\theta)$
		is a periodic function with period $\pi/2$. 
		(c) Angular dependence of Fermi arcs in the range
		of $[0,\pi/4]$, which is between Figures~\ref{fig:periodic_arcs}(b) and (c).
		The parameters are $A=1$, $M_{0}=4.5$, $t=1$, $\epsilon=0.2$,
		and $m=0.5$. \label{fig:periodic_mass}}
\end{figure}

In Figure~\ref{fig:periodic_mass} (a), the Chern number in
$k_{x}k_{y}$-plane against $k_{z}$ is plotted. Here, 
without loss of generality, we assume $k_{\text{w}2}<k_{\text{w}1}$, for the chosen parameters 
given in the caption of Figure~\ref{fig:periodic_mass}. We can see that the
Chern number changes by $2$ when passing a double-Weyl point.
This is because the Chern number on the 
surface that encloses the Weyl point equals to the monopole charge,
as it does for conventional Weyl phases. This explains the non-trivial Chern number
 between the upper pair $k_{z}\in(k_{\text{w}2},k_{\text{w}1})$
and lower pair $k_{z}\in(-k_{\text{w}1},-k_{\text{w}2})$ of Weyl points.
However, the
distribution of the Chern number in the intertwined phase is different from conventional ones. In order to host 
higher-order topological phases, it is required to have a region 
where the Chern number vanishes. Here, the region is between the lower and upper pairs of Weyl points, i.e., $k_z\in\left(-k_{\text{w}2},k_{\text{w}2}\right)$, as shown by the HOT region (green arrow) in Figure~\ref{fig:periodic_mass}. In this region, the existence of boundary states depends on the orientation $\theta$, as shown by  Figure~\ref{fig:periodic_mass} (c). In contrast, in the region with non-zero Chern number, i.e., inside the two sets (upper and lower) of Weyl points, the Fermi arcs exist regardless of the angle $\theta$.

\vspace{0.5cm}
\noindent\textbf{Effective boundary theory}\quad\\
To understand the periodic behavior of Fermi-arc topology, a boundary
theory applicable to any surface orientation is constructive.
We can achieve this goal by firstly deriving two boundary states for
each $\theta$ in the absence of the higher-order term $m\tau_{1}\sigma_{1}$
in Eq.~(\ref{eq:lattice-model}). The spinor part of two boundary states takes the form of
$\psi_{1}\propto(e^{-2i\theta},\sqrt{2}+1,0,0)$
approximately in the region of $k_{z}\in(-k_{\text{w}2},k_{\text{w}2})$,
and $\psi_{2}\propto(0,0,e^{-2i\theta},\sqrt{2}+1)$ in the region
of $k_{z}\in(-k_{\text{w}1},k_{\text{w}1})$. Note that the contribution from spatial part of 
the boundary states does not affect the main results, and is neglected for simplicity.
 By projecting the whole Hamiltonian
into the subspace spanned by the boundary states, we can obtain the
effective boundary Hamiltonian as (See Supplementary Note 1 and 2 for derivation details.)
\begin{equation}
h_{\text{surf}}=-\frac{1}{\sqrt{2}}\left[(2k_{\parallel}^{2}-k_{\text{c}}^{2})\sigma_{3}-m\cos(2\theta)\sigma_{1}\right],\label{eq:lattice-boundary}
\end{equation}
up to a constant term $-1/\sqrt{2}\epsilon\sigma_{0}$, and $k_{\text{c}}^{2}=2t\cos k_{z}-M$.
Clearly, the boundary Hamiltonian is in the form of Eq.~(\ref{eq:boundary-hal}).
The effective Hamiltonian is valid in the region where $\psi_{1}$
and $\psi_{2}$ overlap, i.e., $k_{z}\in(-k_{\text{w}2},k_{\text{w}_{2}})$ between
the middle two Weyl points, where the Chern number is trivial. On a 1D hinge, the gapless points
at $m\cos(2\theta)=0$ result in hinge Fermi arcs, as shown by the green solid lines in Figure~\ref{fig:intertwine}. 

On 2D surfaces, clearly, the periodic change of Fermi-arc topology, shown in Figures~\ref{fig:periodic_arcs} (b)-(d),
is caused by the higher-order term $m\cos(2\theta)$. The period is $\pi/2$, as determined by $m^{2}\cos^{2}(2\theta)$ in the spectrum, in accordance with the general theory of Eq.~(\ref{eq:mass-period}).
Within a single period, the higher-order topology enforces the appearance of
gapless point [Eq.~(\ref{eq:antiperiodic})], which is located at $m\cos(2\theta)=0$. 
The gapless point drastically changes the topology of Fermi arcs, 
because it determines whether the arcs can go through the region of $k_{z}\in(-k_{\text{w}2},k_{\text{w}2})$ between
the middle two Weyl points or not. Figures~\ref{fig:periodic_arcs} (b)-(d) show three representative orientations in a full period
of $\theta\in[0,\pi/2]$. The Fermi arcs cannot go through the region
between the two middle Weyl points at $\theta=0$. After rotating to $\theta=\pi/4$
at the gapless point, the Fermi arcs are allowed
to go through. Finally, at $\theta=\pi/2$, the
Fermi arcs return to their initial topology, completing one period.

\vspace{0.5cm}
\noindent\textbf{Guidelines for material search}\quad\\
Two key ingredients of the discussed intertwined Weyl phases are fourfold (sixfold) rotation symmetry and the unconventional Weyl points on the rotation axis. 
Both requirements can be fulfilled in tetragonal, cubic, or hexagonal space groups, where the desired symmetries emerge. 
There is a correspondence between the absolute value of the chirality of a crossing and the rotation eigenvalues of the involved bands~\cite{composite_Weyl_Vanderbilt}, such that a crossing between certain rotation eigenvalues leads to the required unconventional Weyl points. 
The total phase of eigenvalues accumulated by such crossings for any pair of bands is restricted by the periodicity of Brillouin zone. 
To be more specific, to obtain crossings of the required charge of $\pm 2$ ($\pm 3$), the phase of the eigenvalues of fourfold (sixfold) rotation must change by $\pi$.
By looking at Figure~\ref{fig:intertwine} it is evident that intertwined Weyl points require an even number of such changes of the eigenvalue phase. 
Thus, the necessary crossings can only occur if the total accumulated phase, which is consistent with the periodicity of the Brillouin zone, is equal to $0 \mod 2\pi$, like it is fulfilled in Eq.~\eqref{eq:lattice-model}.
This excludes for nonsymmorphic systems certain filling factors, where the $k$-dependence of rotation eigenvalues would, for example, require an odd number of crossings. 
An appropriate filling $b$ for a $2n$-fold rotation with a fractional translation of $\tfrac{m}{2n}$ has to fulfill $ \tfrac{bm}{2n} \in \mathbb{Z}$.

Commonly, a set of four crossings like in Figure~\ref{fig:intertwine} would be accidental.
Nevertheless, enforced crossings can be found within fourfold degenerate points, e.g., in space groups 106 and 133 on the path M-A.
These comprise unconventional Weyl points related by mirror symmetry \cite{moritz_symmetry_2021}. 
Hereby, mirror pairs of enforced unconventional Weyl points coincide.
If a weak time-reversal and mirror symmetry breaking is introduced to such a system, one may obtain Weyl points in the configuration discussed in Figure~\ref{fig:intertwine}(a) with no other enforced bands at the Fermi energy.
A comprehensive material search is left for future works.

\begin{figure}
	\centering
	\includegraphics[width=0.45\textwidth]{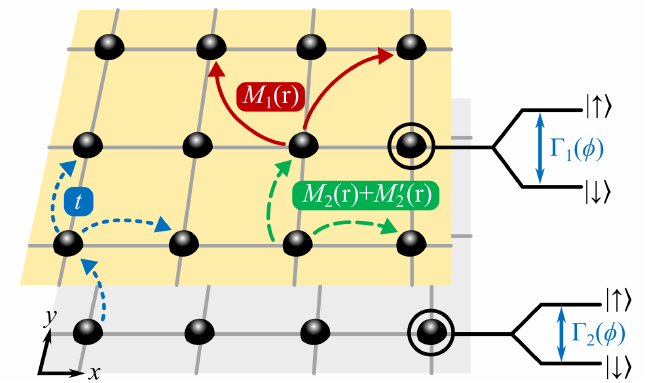}
	\caption{\textbf{Illustration of experimental setup.}
		The atoms are confined in a bilayer optical lattice.
		Couplings between opposite pseudo-spins are processed via laser fields of modes $M_{1}({\bf r})$, $M_{2}({\bf r})$, and $M_{2}'({\bf r})$.
		The on-site energy offset is prepared as $\Gamma_{\lambda}(\phi)$ which not only depends on the layer index but also is manually controlled by the parameter $\phi$.}
	\label{fig:setup}
\end{figure}

\vspace{0.5cm}
\noindent\textbf{Cold-atom experimental realization}\quad\\
Owing to technical advances, cold atoms have been widely applied in quantum simulations of topological matter \cite{Zhang_cold_atom,Wu2016sci,Song2019natphys,Wang2021sci}, and now are also readily available for realizing our intertwined double-Weyl semimetal described by Eq.~(\ref{eq:lattice-model}).
Here we present the realization proposal using fermionic atoms.
We choose two hyperfine states as the pseudo-spins $\uparrow\downarrow$ for the $\sigma$ degrees of freedom.
For our model Hamiltonian, which requires two extra degrees of freedom,
we consider the atomic gases loaded in a 2D bilayer optical lattice,
and thus, the layer index $\lambda$ is used to represent the $\tau$ subspace.
The setup is illustrated in Figure~\ref{fig:setup}.
The energy offset of the hyperfine states is prepared as $\Gamma_\lambda(\phi)=M_0+(-1)^\lambda\epsilon - 2t\cos\phi$
that is manually controllable by the cyclical parameter $\phi$.
It not only generates the on-site energy of the model Hamiltonian,
but also introduces the additional parameterized space represented by $\phi$.
In this way, the band physics of the Weyl Hamiltonian can be captured in the 3D parameterized space $(k_x,k_y,\phi)$ \cite{Zhang2015pra}.
Here, $\phi$ corresponds to $k_z$ in Eq.~(\ref{eq:lattice-model}).

In order to engineer the intra-layer spin-flipped hopping,
we use laser fields of three modes to couple the pseudo-spins. The spatial modulations of the field modes are prepared as
$M_1({\bf r})=iM_1\sin(k_\text{L}x)\sin(k_\text{L}y)$,
$M_2({\bf r})=M_2\cos(k_\text{L}x)\cos(3k_\text{L}y)\cos(k_\text{L}z)$,
and $M_2'({\bf r})=-M_2\cos(3k_\text{L}x)\cos(k_\text{L}y)\cos(k_\text{L}z)$,
where $k_\text{L}=\pi/d$ and $d$ denotes the lattice constant.
Due to the odd parity of $M_1({\bf r})$ in the $xy$-plane \cite{Zheng2019prr},
the on-site and nearest-neighbor (NN) couplings vanish, while the next-NN coupling dominates, resulting in $\sin k_x\sin k_y\sigma_1$.
Due to the crystalline symmetry,
the combination of $M_2({\bf r})$ and $M_2'({\bf r})$ leads to the NN coupling $(\cos k_y-\cos k_x)\sigma_1$.
After making operator transformations,
all the intra-layer terms host opposite signs for different layers. Furthermore, the higher-order topological term $\tau_1\sigma_1$ is naturally introduced by the inter-layer hopping.
The details of the realization proposal are shown in the Supplementary Note 3.

\vspace{0.5cm}
\noindent\textbf{DISCUSSION}\quad\\
Based on the theory to create topological phases by intertwining existing ones, 
we have discovered the intertwined Weyl phase. The intertwined Weyl phase
is different from the individual unconventional Weyl semimetallic phase or higher-order topological phase, and exhibits its distinct characteristic topological features. 
The intertwining results in the drastic change of Fermi-arc topology in a periodic pattern, which constitutes the characteristic feature of the intertwined phase.
 We have proposed a feasible cold-atom experiment to verify our theory and to realize the intertwined Weyl phases.

Our theory could serve as a guiding principle to generate topological phases based on existing ones. A
direct application would be to investigate the intertwining between topological gapless phases with emergent
particles other than Weyl fermions and topological crystalline phases, that are protected by the same symmetry~\citep{po_symmetry-based_2017,tang_comprehensive_2019,zhang_catalogue_2019,hasan_weyl_2021,yu_encyclopedia_2021}.

Finally, we note that Fermi-surface topology is crucial for electronic properties of
material. 
The change of Fermi-arc topology in the intertwined Weyl phases offers a direction of tuning Fermi-surface topology. It would inspire further research on electrical, magnetic, thermodynamic, and transport properties, that are determined by Fermi-surface topology. 
For instance, a potential application is to investigate the quantum oscillations of Fermi-arc surface
states~\cite{potter_quantum_2014}. These oscillations are periodic against the inverse of magnetic field $1/B$. Their frequency $F$ is determined by the Onsager relation $F=\Phi_0/(2\pi^2)A_\text{s}$. Here $\Phi_0=h/2e$ is the magnetic flux quantum, and $A_\text{s}$ is the Fermi surface cross section. In intertwined Weyl phases, the cross section $A_\text{s}$ depends on the surface orientation\ $\theta$. As shown in Figure~\ref{fig:periodic_mass} (c), by changing $\theta$, one may interpolate between small Fermi surfaces (localized around $k_{\text{w}1}$ and $k_{\text{w}2}$) and extended Fermi surfaces (connecting $k_{\text{w}1(2)}$ to $-k_{\text{w}1(2)}$), which must yield a substantial change in the observed quantum oscillation spectra. 
Thus, the intertwined Weyl phases are expected to have a significant change of quantum oscillations upon rotating surface termination.

\quad\\
\vspace{0.3cm}

\noindent
\textbf{METHODS}\\
\textbf{Analytical derivation and symmetry analysis}\\
The derivation of the analytical results makes use of the $k\cdot p$ methods and symmetry analysis. 
The numerical calculation is based on tight-binding model and Green's function method.
Details of these derivations are given in the Supplementary Notes 1, 2, and 4. This includes the 
calculation of Fermi arcs in intertwined double-Weyl semimetals, and the analysis of
symmetry for intertwined double-Weyl and triple-Weyl semimetals. The Supplementary Note 3 
also contains a detailed discussion of the cold-atom realization of intertwined double-Weyl semimetals.
\vspace{0.3cm}

\noindent
\textbf{DATA AVAILABILITY}\\
The numerical data of the plots within this paper are available from the corresponding author upon reasonable request.
\quad\\

\noindent
\textbf{CODE AVAILABILITY}\\
The numerical codes that support the findings of this paper are available from the corresponding author upon reasonable request.
\quad\\

\vspace{0.3cm}

\noindent
\textbf{ACKNOWLEDGMENTS}\\
This work was supported by the Key-Area
Research and Development Program of GuangDong Province (Grant No.
2019B030330001), the CRFs (Nos. C6005-17G and C6009-20G) and GRF (No. 17300220) of Hong Kong. The authors at HKU also thank
support from Guangdong-Hong Kong Joint Laboratory
of Quantum Matter. 

\vspace{0.3cm}

\noindent
\textbf{COMPETING INTERESTS}\\
The authors declare no competing interests.

\vspace{0.3cm}

\noindent
\textbf{AUTHOR CONTRIBUTIONS}\\
W.B.R., C.J.W., and Z.D.W. conceived the project.
W.B.R., M.M.H., and S.B.Z. did the theoretical calculations. 
W.B.R. performed the numerical simulations.
Z.Z. proposed the cold-atom realization, and M.M.H. analyzed the candidate materials.
All authors authored, commented, and corrected the manuscript.


\begin{thebibliography}{10}
	\expandafter\ifx\csname url\endcsname\relax
	\def\url#1{\texttt{#1}}\fi
	\expandafter\ifx\csname urlprefix\endcsname\relax\def\urlprefix{URL }\fi
	\providecommand{\bibinfo}[2]{#2}
	\providecommand{\eprint}[2][]{\url{#2}}
	
	\bibitem{Chiu_RMP_2016}
	\bibinfo{author}{Chiu, C.-K.}, \bibinfo{author}{Teo, J. C.~Y.},
	\bibinfo{author}{Schnyder, A.~P.} \& \bibinfo{author}{Ryu, S.}
	\newblock \bibinfo{title}{Classification of topological quantum matter with
		symmetries}.
	\newblock \emph{\bibinfo{journal}{Rev. Mod. Phys.}}
	\textbf{\bibinfo{volume}{88}}, \bibinfo{pages}{035005}
	(\bibinfo{year}{2016}).
	
	\bibitem{Armitage2018}
	\bibinfo{author}{Armitage, N.~P.}, \bibinfo{author}{Mele, E.~J.} \&
	\bibinfo{author}{Vishwanath, A.}
	\newblock \bibinfo{title}{Weyl and Dirac semimetals in three-dimensional
		solids}.
	\newblock \emph{\bibinfo{journal}{Rev. Mod. Phys.}}
	\textbf{\bibinfo{volume}{90}}, \bibinfo{pages}{015001}
	(\bibinfo{year}{2018}).
	
	\bibitem{Yang_PRL_2014}
	\bibinfo{author}{Yang, S.~A.}, \bibinfo{author}{Pan, H.} \&
	\bibinfo{author}{Zhang, F.}
	\newblock \bibinfo{title}{Dirac and Weyl superconductors in three dimensions}.
	\newblock \emph{\bibinfo{journal}{Phys. Rev. Lett.}}
	\textbf{\bibinfo{volume}{113}}, \bibinfo{pages}{046401}
	(\bibinfo{year}{2014}).
	
	\bibitem{Wan_PRB_2011}
	\bibinfo{author}{Wan, X.}, \bibinfo{author}{Turner, A.~M.},
	\bibinfo{author}{Vishwanath, A.} \& \bibinfo{author}{Savrasov, S.~Y.}
	\newblock \bibinfo{title}{Topological semimetal and Fermi-arc surface states in
		the electronic structure of pyrochlore iridates}.
	\newblock \emph{\bibinfo{journal}{Phys. Rev. B}} \textbf{\bibinfo{volume}{83}},
	\bibinfo{pages}{205101} (\bibinfo{year}{2011}).
	
	\bibitem{xu_discovery_2015}
	\bibinfo{author}{Xu, S.-Y.} \emph{et~al.}
	\newblock \bibinfo{title}{Discovery of a Weyl fermion semimetal and topological
		Fermi arcs}.
	\newblock \emph{\bibinfo{journal}{Science}} \textbf{\bibinfo{volume}{349}},
	\bibinfo{pages}{613--617} (\bibinfo{year}{2015}).
	
	\bibitem{Lv_Weyl_2015}
	\bibinfo{author}{Lv, B.~Q.} \emph{et~al.}
	\newblock \bibinfo{title}{Experimental discovery of Weyl semimetal TaAs}.
	\newblock \emph{\bibinfo{journal}{Phys. Rev. X}} \textbf{\bibinfo{volume}{5}},
	\bibinfo{pages}{031013} (\bibinfo{year}{2015}).
	
	\bibitem{Meng_PRB_2012}
	\bibinfo{author}{Meng, T.} \& \bibinfo{author}{Balents, L.}
	\newblock \bibinfo{title}{Weyl superconductors}.
	\newblock \emph{\bibinfo{journal}{Phys. Rev. B}} \textbf{\bibinfo{volume}{86}},
	\bibinfo{pages}{054504} (\bibinfo{year}{2012})
	
	\bibitem{Morali_magnetic_Weyl_2019}
	\bibinfo{author}{Morali, N.} \emph{et~al.}
	\newblock \bibinfo{title}{Fermi-arc diversity on surface terminations of the
		magnetic Weyl semimetal Co$_3$Sn$_2$S$_2$}.
	\newblock \emph{\bibinfo{journal}{Science}} \textbf{\bibinfo{volume}{365}},
	\bibinfo{pages}{1286--1291} (\bibinfo{year}{2019}).
	
	\bibitem{Liu_magnetic_Weyl_2019}
	\bibinfo{author}{Liu, D.~F.} \emph{et~al.}
	\newblock \bibinfo{title}{Magnetic Weyl semimetal phase in a kagom{\'e}
		crystal}.
	\newblock \emph{\bibinfo{journal}{Science}} \textbf{\bibinfo{volume}{365}},
	\bibinfo{pages}{1282--1285} (\bibinfo{year}{2019}).
	
	\bibitem{Bradlynaaf5037}
	\bibinfo{author}{Bradlyn, B.} \emph{et~al.}
	\newblock \bibinfo{title}{Beyond Dirac and Weyl fermions: Unconventional
		quasiparticles in conventional crystals}.
	\newblock \emph{\bibinfo{journal}{Science}} \textbf{\bibinfo{volume}{353}},
	\bibinfo{pages}{aaf5037} (\bibinfo{year}{2016}).
	
	\bibitem{sanchez_topological_2019}
	\bibinfo{author}{Sanchez, D.~S.} \emph{et~al.}
	\newblock \bibinfo{title}{Topological chiral crystals with helicoid-arc quantum
		states}.
	\newblock \emph{\bibinfo{journal}{Nature}} \textbf{\bibinfo{volume}{567}},
	\bibinfo{pages}{500--505} (\bibinfo{year}{2019}).
	
	\bibitem{yang_topological_2019}
	\bibinfo{author}{Yang, Y.} \emph{et~al.}
	\newblock \bibinfo{title}{Topological triply degenerate point with double
		{Fermi} arcs}.
	\newblock \emph{\bibinfo{journal}{Nat. Phys.}}
	\textbf{\bibinfo{volume}{15}}, \bibinfo{pages}{645--649}
	(\bibinfo{year}{2019}).
	
	\bibitem{he_observation_2020}
	\bibinfo{author}{He, H.} \emph{et~al.}
	\newblock \bibinfo{title}{Observation of quadratic {Weyl} points and
		double-helicoid arcs}.
	\newblock \emph{\bibinfo{journal}{Nat. Commun.}}
	\textbf{\bibinfo{volume}{11}}, \bibinfo{pages}{1820} (\bibinfo{year}{2020}).
	
	
	\bibitem{Fang2012PRL}
	\bibinfo{author}{Fang, C.}, \bibinfo{author}{Gilbert, M.~J.},
	\bibinfo{author}{Dai, X.} \& \bibinfo{author}{Bernevig, B.~A.}
	\newblock \bibinfo{title}{Multi-Weyl topological semimetals stabilized by point
		group symmetry}.
	\newblock \emph{\bibinfo{journal}{Phys. Rev. Lett.}}
	\textbf{\bibinfo{volume}{108}}, \bibinfo{pages}{266802}
	(\bibinfo{year}{2012}).
	
	\bibitem{chen_photonic_2016}
	\bibinfo{author}{Chen, W.-J.}, \bibinfo{author}{Xiao, M.} \&
	\bibinfo{author}{Chan, C.~T.}
	\newblock \bibinfo{title}{Photonic crystals possessing multiple {Weyl} points
		and the experimental observation of robust surface states}.
	\newblock \emph{\bibinfo{journal}{Nat. Commun.}}
	\textbf{\bibinfo{volume}{7}}, \bibinfo{pages}{13038} (\bibinfo{year}{2016}).
	
	\bibitem{huang_new_2016}
	\bibinfo{author}{Huang, S.-M.} \emph{et~al.}
	\newblock \bibinfo{title}{New type of {Weyl} semimetal with quadratic double
		{Weyl} fermions}.
	\newblock \emph{\bibinfo{journal}{Proc. Natl. Acad. Sci. U.S.A.}} \textbf{\bibinfo{volume}{113}}, \bibinfo{pages}{1180--1185}
	(\bibinfo{year}{2016}).
	
	\bibitem{Vaidya_2020_charge2}
	\bibinfo{author}{Vaidya, S.} \emph{et~al.}
	\newblock \bibinfo{title}{Observation of a charge-2 photonic Weyl point in the
		infrared}.
	\newblock \emph{\bibinfo{journal}{Phys. Rev. Lett.}}
	\textbf{\bibinfo{volume}{125}}, \bibinfo{pages}{253902}
	(\bibinfo{year}{2020}).
	
	\bibitem{Yang2020PRL}
	\bibinfo{author}{Yang, Y.} \emph{et~al.}
	\newblock \bibinfo{title}{Ideal unconventional Weyl point in a chiral photonic
		metamaterial}.
	\newblock \emph{\bibinfo{journal}{Phys. Rev. Lett.}}
	\textbf{\bibinfo{volume}{125}}, \bibinfo{pages}{143001}
	(\bibinfo{year}{2020}).
	
	\bibitem{Dantas2020PRR}
	\bibinfo{author}{Dantas, R. M.~A.}, \bibinfo{author}{Pe\~na Benitez, F.},
	\bibinfo{author}{Roy, B.} \& \bibinfo{author}{Sur\'owka, P.}
	\newblock \bibinfo{title}{Non-abelian anomalies in multi-Weyl semimetals}.
	\newblock \emph{\bibinfo{journal}{Phys. Rev. Res.}}
	\textbf{\bibinfo{volume}{2}}, \bibinfo{pages}{013007} (\bibinfo{year}{2020}).
	
	\bibitem{Liang2011crystalline}
	\bibinfo{author}{Fu, L.}
	\newblock \bibinfo{title}{Topological crystalline insulators}.
	\newblock \emph{\bibinfo{journal}{Phys. Rev. Lett.}}
	\textbf{\bibinfo{volume}{106}}, \bibinfo{pages}{106802}
	(\bibinfo{year}{2011}).
	
	
	\bibitem{hsieh_topological_2012}
	\bibinfo{author}{Hsieh, T.~H.} \emph{et~al.}
	\newblock \bibinfo{title}{Topological crystalline insulators in the {SnTe}
		material class}.
	\newblock \emph{\bibinfo{journal}{Nat. Commun.}}
	\textbf{\bibinfo{volume}{3}}, \bibinfo{pages}{982} (\bibinfo{year}{2012}).
	
	\bibitem{tanaka_experimental_2012}
	\bibinfo{author}{Tanaka, Y.} \emph{et~al.}
	\newblock \bibinfo{title}{Experimental realization of a topological crystalline
		insulator in {SnTe}}.
	\newblock \emph{\bibinfo{journal}{Nat. Phys.}}
	\textbf{\bibinfo{volume}{8}}, \bibinfo{pages}{800--803}
	(\bibinfo{year}{2012}).
	
	\bibitem{dziawa_topological_2012}
	\bibinfo{author}{Dziawa, P.} \emph{et~al.}
	\newblock \bibinfo{title}{Topological crystalline insulator states in PbSnSe}.
	\newblock \emph{\bibinfo{journal}{Nat. Mater.}}
	\textbf{\bibinfo{volume}{11}}, \bibinfo{pages}{1023--1027}
	(\bibinfo{year}{2012}).
	
	\bibitem{Ando_Topological_2015}
	\bibinfo{author}{Ando, Y.} \& \bibinfo{author}{Fu, L.}
	\newblock \bibinfo{title}{Topological crystalline insulators and topological
		superconductors: From concepts to materials}.
	\newblock \emph{\bibinfo{journal}{Annu. Rev. Condens. Matter Phys.}}
	\textbf{\bibinfo{volume}{6}}, \bibinfo{pages}{361--381}
	(\bibinfo{year}{2015}).
	
	\bibitem{Fang_2015_new}
	\bibinfo{author}{Fang, C.} \& \bibinfo{author}{Fu, L.}
	\newblock \bibinfo{title}{New classes of three-dimensional topological
		crystalline insulators: Nonsymmorphic and magnetic}.
	\newblock \emph{\bibinfo{journal}{Phys. Rev. B}} \textbf{\bibinfo{volume}{91}},
	\bibinfo{pages}{161105} (\bibinfo{year}{2015}).
	
	\bibitem{Slager_RPX_2017}
	\bibinfo{author}{Kruthoff, J.} \emph{et~al.}
	\newblock \bibinfo{title}{Topological classification of crystalline insulators
		through band structure combinatorics}.
	\newblock \emph{\bibinfo{journal}{Phys. Rev. X}} \textbf{\bibinfo{volume}{7}},
	\bibinfo{pages}{041069} (\bibinfo{year}{2017}).
	
	\bibitem{Khalaf_2018_symmetry}
	\bibinfo{author}{Khalaf, E.}, \bibinfo{author}{Po, H.~C.},
	\bibinfo{author}{Vishwanath, A.} \& \bibinfo{author}{Watanabe, H.}
	\newblock \bibinfo{title}{Symmetry indicators and anomalous surface states of
		topological crystalline insulators}.
	\newblock \emph{\bibinfo{journal}{Phys. Rev. X}} \textbf{\bibinfo{volume}{8}},
	\bibinfo{pages}{031070} (\bibinfo{year}{2018}).
	
	\bibitem{po_symmetry-based_2017}
	\bibinfo{author}{Po, H.~C.}, \bibinfo{author}{Vishwanath, A.} \&
	\bibinfo{author}{Watanabe, H.}
	\newblock \bibinfo{title}{Symmetry-based indicators of band topology in the 230
		space groups}.
	\newblock \emph{\bibinfo{journal}{Nat. Commun.}}
	\textbf{\bibinfo{volume}{8}}, \bibinfo{pages}{50} (\bibinfo{year}{2017}).
	
	\bibitem{tang_comprehensive_2019}
	\bibinfo{author}{Tang, F.}, \bibinfo{author}{Po, H.~C.},
	\bibinfo{author}{Vishwanath, A.} \& \bibinfo{author}{Wan, X.}
	\newblock \bibinfo{title}{Comprehensive search for topological materials using
		symmetry indicators}.
	\newblock \emph{\bibinfo{journal}{Nature}} \textbf{\bibinfo{volume}{566}},
	\bibinfo{pages}{486--489} (\bibinfo{year}{2019}).
	
	\bibitem{zhang_catalogue_2019}
	\bibinfo{author}{Zhang, T.} \emph{et~al.}
	\newblock \bibinfo{title}{Catalogue of topological electronic materials}.
	\newblock \emph{\bibinfo{journal}{Nature}} \textbf{\bibinfo{volume}{566}},
	\bibinfo{pages}{475--479} (\bibinfo{year}{2019}).
	
	\bibitem{vergniory_complete_2019}
	\bibinfo{author}{Vergniory, M.~G.} \emph{et~al.}
	\newblock \bibinfo{title}{A complete catalogue of high-quality topological
		materials}.
	\newblock \emph{\bibinfo{journal}{Nature}} \textbf{\bibinfo{volume}{566}},
	\bibinfo{pages}{480--485} (\bibinfo{year}{2019}).
	
	\bibitem{Benalcazar61}
	\bibinfo{author}{Benalcazar, W.~A.}, \bibinfo{author}{Bernevig, B.~A.} \&
	\bibinfo{author}{Hughes, T.~L.}
	\newblock \bibinfo{title}{Quantized electric multipole insulators}.
	\newblock \emph{\bibinfo{journal}{Science}} \textbf{\bibinfo{volume}{357}},
	\bibinfo{pages}{61--66} (\bibinfo{year}{2017}).
	
	\bibitem{song_PRL_2017}
	\bibinfo{author}{Song, Z.}, \bibinfo{author}{Fang, Z.} \&
	\bibinfo{author}{Fang, C.}
	\newblock \bibinfo{title}{$(d\ensuremath{-}2)$-dimensional edge states of
		rotation symmetry protected topological states}.
	\newblock \emph{\bibinfo{journal}{Phys. Rev. Lett.}}
	\textbf{\bibinfo{volume}{119}}, \bibinfo{pages}{246402}
	(\bibinfo{year}{2017}).
	
	\bibitem{Piet_PRL_2017}
	\bibinfo{author}{Langbehn, J.}  \emph{et~al.}
	\newblock \bibinfo{title}{Reflection-symmetric second-order topological
		insulators and superconductors}.
	\newblock \emph{\bibinfo{journal}{Phys. Rev. Lett.}}
	\textbf{\bibinfo{volume}{119}}, \bibinfo{pages}{246401}
	(\bibinfo{year}{2017}).
	
	
	\bibitem{Peterson_fractional_2020}
	\bibinfo{author}{Peterson, C.~W.}  \emph{et~al.}
	\newblock \bibinfo{title}{A fractional corner anomaly reveals higher-order
		topology}.
	\newblock \emph{\bibinfo{journal}{Science}} \textbf{\bibinfo{volume}{368}},
	\bibinfo{pages}{1114--1118} (\bibinfo{year}{2020}).
	
	\bibitem{Schindl18science}
	\bibinfo{author}{Schindler, F.} \emph{et~al.}
	\newblock \bibinfo{title}{Higher-order topological insulators}.
	\newblock \emph{\bibinfo{journal}{Sci. Adv.}} \textbf{\bibinfo{volume}{4}},
	\bibinfo{pages}{eaat0346} (\bibinfo{year}{2018}).
	
	\bibitem{Schindler2018higher}
	\bibinfo{author}{Schindler, F.} \emph{et~al.}
	\newblock \bibinfo{title}{Higher-order topology in bismuth}.
	\newblock \emph{\bibinfo{journal}{Nat. Phys.}}
	\textbf{\bibinfo{volume}{14}}, \bibinfo{pages}{918--924}
	(\bibinfo{year}{2018}).
	
	\bibitem{serra2018observation}
	\bibinfo{author}{Serra-Garcia, M.} \emph{et~al.}
	\newblock \bibinfo{title}{Observation of a phononic quadrupole topological
		insulator}.
	\newblock \emph{\bibinfo{journal}{Nature}} \textbf{\bibinfo{volume}{555}},
	\bibinfo{pages}{342--345} (\bibinfo{year}{2018}).
	
	\bibitem{Ezawa18PRL}
	\bibinfo{author}{Ezawa, M.}
	\newblock \bibinfo{title}{Higher-order topological insulators and semimetals on
		the breathing kagome and pyrochlore lattices}.
	\newblock \emph{\bibinfo{journal}{Phys. Rev. Lett.}}
	\textbf{\bibinfo{volume}{120}}, \bibinfo{pages}{026801}
	(\bibinfo{year}{2018}).
	
	\bibitem{Bernevig2019PRL}
	\bibinfo{author}{Wang, Z.} \emph{et~al.}
	\newblock \bibinfo{title}{Higher-order topology, monopole nodal lines, and the
		origin of large fermi arcs in transition metal dichalcogenides
		$X{\mathrm{Te}}_{2}$ ($X=\mathrm{Mo},\mathrm{W}$)}.
	\newblock \emph{\bibinfo{journal}{Phys. Rev. Lett.}}
	\textbf{\bibinfo{volume}{123}}, \bibinfo{pages}{186401}
	(\bibinfo{year}{2019}).
	
	\bibitem{ChenR20PRL}
	\bibinfo{author}{Chen, R.}  \emph{et~al.}
	\newblock \bibinfo{title}{Higher-order topological insulators in
		quasicrystals}.
	\newblock \emph{\bibinfo{journal}{Phys. Rev. Lett.}}
	\textbf{\bibinfo{volume}{124}}, \bibinfo{pages}{036803}
	(\bibinfo{year}{2020}).
	
	\bibitem{ZhangSB20arxiv2}
	\bibinfo{author}{{Zhang}, S.-B.}, \bibinfo{author}{{Calzona}, A.} \&
	\bibinfo{author}{{Trauzettel}, B.}
	\newblock \bibinfo{title}{{All-electrically tunable networks of Majorana bound
			states}}.
	\newblock \emph{\bibinfo{journal}{Phys. Rev. B}}
	\textbf{\bibinfo{volume}{102}}, \bibinfo{pages}{100503}
	(\bibinfo{year}{2020}).
	
	\bibitem{Trifunovic19PRX}
	\bibinfo{author}{Trifunovic, L.} \& \bibinfo{author}{Brouwer, P.~W.}
	\newblock \bibinfo{title}{Higher-order bulk-boundary correspondence for
		topological crystalline phases}.
	\newblock \emph{\bibinfo{journal}{Phys. Rev. X}} \textbf{\bibinfo{volume}{9}},
	\bibinfo{pages}{011012} (\bibinfo{year}{2019}).
	
	\bibitem{Eslam_PRB_2018}
	\bibinfo{author}{Khalaf, E.}
	\newblock \bibinfo{title}{Higher-order topological insulators and
		superconductors protected by inversion symmetry}.
	\newblock \emph{\bibinfo{journal}{Phys. Rev. B}} \textbf{\bibinfo{volume}{97}},
	\bibinfo{pages}{205136} (\bibinfo{year}{2018}).
	
	\bibitem{ZhangSB20quantumcomputation}
	\bibinfo{author}{Zhang, S.-B.} \emph{et~al.}
	\newblock \bibinfo{title}{Topological and holonomic quantum computation based
		on second-order topological superconductors}.
	\newblock \emph{\bibinfo{journal}{Phys. Rev. Res.}}
	\textbf{\bibinfo{volume}{2}}, \bibinfo{pages}{043025} (\bibinfo{year}{2020}).
	
	\bibitem{Lin_PRB_2018}
	\bibinfo{author}{Lin, M.} \& \bibinfo{author}{Hughes, T.~L.}
	\newblock \bibinfo{title}{Topological quadrupolar semimetals}.
	\newblock \emph{\bibinfo{journal}{Phys. Rev. B}} \textbf{\bibinfo{volume}{98}},
	\bibinfo{pages}{241103} (\bibinfo{year}{2018}).
	
	\bibitem{wang2020boundary}
	\bibinfo{author}{Wang, K.} \emph{et~al.}
	\newblock \bibinfo{title}{Boundary criticality of $\mathcal{PT}$-invariant
		topology and second-order nodal-line semimetals}.
	\newblock \emph{\bibinfo{journal}{Phys. Rev. Lett.}}
	\textbf{\bibinfo{volume}{125}}, \bibinfo{pages}{126403}
	(\bibinfo{year}{2020}).
	
	\bibitem{ghorashi_second-order_2019}
	\bibinfo{author}{Ghorashi, S. A.~A.}, \bibinfo{author}{Hu, X.},
	\bibinfo{author}{Hughes, T.~L.} \& \bibinfo{author}{Rossi, E.}
	\newblock \bibinfo{title}{Second-order {Dirac} superconductors and magnetic
		field induced Majorana hinge modes}.
	\newblock \emph{\bibinfo{journal}{Phys. Rev. B}}
	\textbf{\bibinfo{volume}{100}}, \bibinfo{pages}{020509}
	(\bibinfo{year}{2019}).
	
	\bibitem{Tiwari2020PRR}
	\bibinfo{author}{Tiwari, A.}, \bibinfo{author}{Jahin, A.} \&
	\bibinfo{author}{Wang, Y.}
	\newblock \bibinfo{title}{Chiral Dirac superconductors: Second-order and
		boundary-obstructed topology}.
	\newblock \emph{\bibinfo{journal}{Phys. Rev. Res.}}
	\textbf{\bibinfo{volume}{2}}, \bibinfo{pages}{043300} (\bibinfo{year}{2020}).
	
	\bibitem{Bitan2019PRR}
	\bibinfo{author}{Roy, B.}
	\newblock \bibinfo{title}{Antiunitary symmetry protected higher-order
		topological phases}.
	\newblock \emph{\bibinfo{journal}{Phys. Rev. Res.}}
	\textbf{\bibinfo{volume}{1}}, \bibinfo{pages}{032048} (\bibinfo{year}{2019}).
	
	\bibitem{wang_2020_higherorder}
	\bibinfo{author}{Wang, H.-X.}  \emph{et~al.}
	\newblock \bibinfo{title}{Higher-order Weyl semimetals}.
	\newblock \emph{\bibinfo{journal}{Phys. Rev. Lett.}}
	\textbf{\bibinfo{volume}{125}}, \bibinfo{pages}{146401}
	(\bibinfo{year}{2020}).
	
	\bibitem{ghorashi_2020_higherorder}
	\bibinfo{author}{Ghorashi, S. A.~A.}, \bibinfo{author}{Li, T.} \&
	\bibinfo{author}{Hughes, T.~L.}
	\newblock \bibinfo{title}{Higher-order Weyl semimetals}.
	\newblock \emph{\bibinfo{journal}{Phys. Rev. Lett.}}
	\textbf{\bibinfo{volume}{125}}, \bibinfo{pages}{266804}
	(\bibinfo{year}{2020}).
	
	\bibitem{rui_higher-order_2020}
	\bibinfo{author}{Rui, W.~B.} \emph{et~al.}
	\newblock \bibinfo{title}{Higher-order Weyl superconductors with anisotropic
		Weyl-point connectivity}.
	\newblock \emph{\bibinfo{journal}{Phys. Rev. B}}
	\textbf{\bibinfo{volume}{103}}, \bibinfo{pages}{184510}
	(\bibinfo{year}{2021}).
	
	\bibitem{wei_higher-order_2021}
	\bibinfo{author}{Wei, Q.} \emph{et~al.}
	\newblock \bibinfo{title}{Higher-order topological semimetal in acoustic
		crystals}.
	\newblock \emph{\bibinfo{journal}{Nat. Mater.}}
	\textbf{\bibinfo{volume}{20}}, \bibinfo{pages}{812–817}
	(\bibinfo{year}{2021}).

	
	\bibitem{luo_observation_2021}
	\bibinfo{author}{Luo, L.} \emph{et~al.}
	\newblock \bibinfo{title}{Observation of a phononic higher-order Weyl
		semimetal}.
	\newblock \emph{\bibinfo{journal}{Nat. Mater.}} 
	\textbf{\bibinfo{volume}{20}}, \bibinfo{pages}{794–799}
	 (\bibinfo{year}{2021}).
	
	\bibitem{zhang_2019_higherorder}
	\bibinfo{author}{Zhang, R.-X.}, \bibinfo{author}{Hsu, Y.-T.} \&
	\bibinfo{author}{Das~Sarma, S.}
	\newblock \bibinfo{title}{Higher-order topological Dirac superconductors}.
	\newblock \emph{\bibinfo{journal}{Phys. Rev. B}}
	\textbf{\bibinfo{volume}{102}}, \bibinfo{pages}{094503}
	(\bibinfo{year}{2020}).
	
	\bibitem{Bitan-general-principle}
	\bibinfo{author}{C\ifmmode \u{a}\else \u{a}\fi{}lug\ifmmode~\u{a}\else
		\u{a}\fi{}ru, D.}, \bibinfo{author}{Juri\ifmmode \check{c}\else
		\v{c}\fi{}i\ifmmode~\acute{c}\else \'{c}\fi{}, V.} \& \bibinfo{author}{Roy,
		B.}
	\newblock \bibinfo{title}{Higher-order topological phases: A general principle
		of construction}.
	\newblock \emph{\bibinfo{journal}{Phys. Rev. B}} \textbf{\bibinfo{volume}{99}},
	\bibinfo{pages}{041301} (\bibinfo{year}{2019}).
	
	\bibitem{Bitan-dirty-higher-order}
	\bibinfo{author}{Szab\'o, A.~L.} \& \bibinfo{author}{Roy, B.}
	\newblock \bibinfo{title}{Dirty higher-order Dirac semimetal: Quantum
		criticality and bulk-boundary correspondence}.
	\newblock \emph{\bibinfo{journal}{Phys. Rev. Res.}}
	\textbf{\bibinfo{volume}{2}}, \bibinfo{pages}{043197} (\bibinfo{year}{2020}).
	
	\bibitem{zhang_second-order_2019}
	\bibinfo{author}{Zhang, X.} \emph{et~al.}
	\newblock \bibinfo{title}{Second-order topology and multidimensional
		topological transitions in sonic crystals}.
	\newblock \emph{\bibinfo{journal}{Nat. Phys.}}
	\textbf{\bibinfo{volume}{15}}, \bibinfo{pages}{582--588}
	(\bibinfo{year}{2019}).
	
	\bibitem{Bitan-dirty-weyl}
	\bibinfo{author}{Bera, S.}, \bibinfo{author}{Sau, J.~D.} \&
	\bibinfo{author}{Roy, B.}
	\newblock \bibinfo{title}{Dirty Weyl semimetals: Stability, phase transition,
		and quantum criticality}.
	\newblock \emph{\bibinfo{journal}{Phys. Rev. B}} \textbf{\bibinfo{volume}{93}},
	\bibinfo{pages}{201302} (\bibinfo{year}{2016}).
	
	\bibitem{Bitan_nonabelian_PRR}
	\bibinfo{author}{Dantas, R. M.~A.}, \bibinfo{author}{Pe\~na Benitez, F.},
	\bibinfo{author}{Roy, B.} \& \bibinfo{author}{Sur\'owka, P.}
	\newblock \bibinfo{title}{Non-abelian anomalies in multi-Weyl semimetals}.
	\newblock \emph{\bibinfo{journal}{Phys. Rev. Res.}}
	\textbf{\bibinfo{volume}{2}}, \bibinfo{pages}{013007} (\bibinfo{year}{2020}).
	
	\bibitem{composite_Weyl_Vanderbilt}
	\bibinfo{author}{Tsirkin, S.~S.}, \bibinfo{author}{Souza, I.} \&
	\bibinfo{author}{Vanderbilt, D.}
	\newblock \bibinfo{title}{Composite Weyl nodes stabilized by screw symmetry
		with and without time-reversal invariance}.
	\newblock \emph{\bibinfo{journal}{Phys. Rev. B}} \textbf{\bibinfo{volume}{96}},
	\bibinfo{pages}{045102} (\bibinfo{year}{2017}).
	
	\bibitem{moritz_symmetry_2021}
	\bibinfo{author}{Hirschmann, M.~M.}\emph{et~al.}
	\newblock \bibinfo{title}{Symmetry-enforced band crossings in tetragonal
		materials: Dirac and Weyl degeneracies on points, lines, and planes}.
	\newblock \emph{\bibinfo{journal}{Phys. Rev. Mater.}}
	\textbf{\bibinfo{volume}{5}}, \bibinfo{pages}{054202} (\bibinfo{year}{2021}).
	
	\bibitem{Zhang_cold_atom}
	\bibinfo{author}{Zhang, D.-W.} \emph{et~al.}
	\newblock \bibinfo{title}{Topological quantum matter with cold atoms}.
	\newblock \emph{\bibinfo{journal}{Adv. Phys.}}
	\textbf{\bibinfo{volume}{67}}, \bibinfo{pages}{253--402}
	(\bibinfo{year}{2018}).
	
	\bibitem{Wu2016sci}
	\bibinfo{author}{Wu, Z.} \emph{et~al.}
	\newblock \bibinfo{title}{{Realization of two-dimensional spin-orbit coupling
			for Bose-Einstein condensates}}.
	\newblock \emph{\bibinfo{journal}{Science}} \textbf{\bibinfo{volume}{354}},
	\bibinfo{pages}{83--88} (\bibinfo{year}{2016}).
	
	\bibitem{Song2019natphys}
	\bibinfo{author}{Song, B.} \emph{et~al.}
	\newblock \bibinfo{title}{{Observation of nodal-line semimetal with ultracold
			fermions in an optical lattice}}.
	\newblock \emph{\bibinfo{journal}{Nat. Phys.}} \textbf{\bibinfo{volume}{15}},
	\bibinfo{pages}{911--916} (\bibinfo{year}{2019}).
	
	\bibitem{Wang2021sci}
	\bibinfo{author}{Wang, Z.-Y.} \emph{et~al.}
	\newblock \bibinfo{title}{{Realization of an ideal Weyl semimetal band in a
			quantum gas with 3D spin-orbit coupling}}.
	\newblock \emph{\bibinfo{journal}{Science}} \textbf{\bibinfo{volume}{372}},
	\bibinfo{pages}{271--276} (\bibinfo{year}{2021}).
	
	
	\bibitem{Zhang2015pra}
	\bibinfo{author}{Zhang, D.-W.}, \bibinfo{author}{Zhu, S.-L.} \&
	\bibinfo{author}{Wang, Z.~D.}
	\newblock \bibinfo{title}{{Simulating and exploring Weyl semimetal physics with
			cold atoms in a two-dimensional optical lattice}}.
	\newblock \emph{\bibinfo{journal}{Phys. Rev. A}} \textbf{\bibinfo{volume}{92}},
	\bibinfo{pages}{013632} (\bibinfo{year}{2015}).
	
	\bibitem{Zheng2019prr}
	\bibinfo{author}{Zheng, Z.}  \emph{et~al.}
	\newblock \bibinfo{title}{{Chiral magnetic effect in three-dimensional optical
			lattices}}.
	\newblock \emph{\bibinfo{journal}{Phys. Rev. Res.}}
	\textbf{\bibinfo{volume}{1}}, \bibinfo{pages}{033102} (\bibinfo{year}{2019}).
	
	\bibitem{hasan_weyl_2021}
	\bibinfo{author}{Hasan, M.~Z.} \emph{et~al.}
	\newblock \bibinfo{title}{Weyl, Dirac and high-fold chiral fermions in
		topological quantum matter}.
	\newblock \emph{\bibinfo{journal}{Nat. Rev. Mater.}}  \textbf{\bibinfo{volume}{6}}, \bibinfo{pages}{784–803}
	(\bibinfo{year}{2021}).
	
	\bibitem{yu_encyclopedia_2021}
	\bibinfo{author}{Yu, Z.-M.} \emph{et~al.}
	\newblock \bibinfo{title}{Encyclopedia of emergent particles in
		three-dimensional crystals}.
	\newblock \bibinfo{journal}{Sci. Bull.}.
	\newblock \bibinfo{doi}{\url{https://doi.org/10.1016/j.scib.2021.10.023}}
	(\bibinfo{year}{2021}).
	
	\bibitem{potter_quantum_2014}
	\bibinfo{author}{Potter, A.~C.}, \bibinfo{author}{Kimchi, I.} \&
	\bibinfo{author}{Vishwanath, A.}
	\newblock \bibinfo{title}{Quantum oscillations from surface Fermi arcs in
		Weyl and Dirac semimetals}.
	\newblock \emph{\bibinfo{journal}{Nat. Commun.}}
	\textbf{\bibinfo{volume}{5}}, \bibinfo{pages}{5161} (\bibinfo{year}{2014}).
	
\end{thebibliography}
\end{document}